# Situation Awareness and Information Fusion in Sales and Customer Engagement: A Paradigm Shift

Yifei Huang
Outreach Corporation
Seattle, USA
yifei.huang@outreach.io

*Abstract*—With today's savvy and empowered customers, sales requires more judgment and becomes more cognitively intense than ever before. We argue that Situation Awareness (SA) is at the center of effective sales and customer engagement in this new era, and Information Fusion (IF) is the key for developing the next generation of decision support systems for digital and AI transformation, leveraging the ubiquitous virtual presence of sales and customer engagement which provides substantially richer capacity to access information. We propose a vision and path for the paradigm shift from Customer Relationship Management (CRM) to the new paradigm of IF. We argue this new paradigm solves major problems of the current CRM paradigm: (1) it reduces the burden of manual data entry and enables more reliable, comprehensive and up-to-date data and knowledge, (2) it enhances individual and team SA and alleviates information silos with increased knowledge transferability, and (3) it enables a more powerful ecosystem of applications by providing common shared layer of computable knowledge assets.

## I. Introduction

Sales is an enormously important function in a market economy. Through salespeople's engagement with potential buyers, innovations of products and services are spread across the market and adopted by customers, generating revenue for further growth and innovations. In the U.S., 15.8 million people were employed in a sales or related occupation in 2018, which represents 10.1% of the employed population [1]. According to Gartner, worldwide spending in Customer Relationship Management (CRM) software reached $48.2 billion in 2018, representing 25% of the entire $193.6 billion worldwide enterprise application software revenue [2].

Three recent trends in sales and customer engagement calls for a paradigm shift. First, as customers are empowered with better information, more choices, and richer resources, the bar for salespeople has been raised. "Sales today requires more judgment than ever before, the cognitive burden on the salesperson is significantly higher – emotional intelligence is not enough." [3] In fact, according to a recent global survey, only 42% of salespeople expect to hit their quota, which was interpreted by the study as "sales team falling short of rising customer expectations" [4]. Second, increasingly ubiquitous virtual presence of sales and customer engagement enables substantially richer capacity to access information [5]. Third, there are tremendous opportunities and buzz about leveraging artificial intelligence (AI) to transform sales and customer engagement [6], which requires a more scalable and sustainable foundation of data and knowledge.

Situation Awareness (SA) is well recognized by human factor studies as the cornerstone of successful decision-making in a dynamic and complex environment [7]. SA is the mental state of a person's dynamic understanding of "what is going on": "the perception of elements in the environment within a volume of time and space, the comprehension of their meaning, and the projection of their status in the near future", as classically defined by Endsley [8]. Information Fusion (IF) aims to achieve machine-assisted SA via "semi-automation of the functionalities of sensation, perception, cognition, comprehension, and projection that is otherwise performed by people" – through engineering systems that enable "association, correlation and combination of data and information from single and multiple sources" to "assemble a representation of aspects of interest in an environment" [9]–[11]. While SA and IF has far-reaching and impactful applications in fields such as aviation, military operations, nuclear power plant operations, and cybersecurity [7], their application in sales is a gold mine not yet explored.

In this paper, we argue that SA is at the center of effective sales and customer engagement in the new era, and IF is the key for developing the next generation of decision support systems for digital and AI transformation. We propose a paradigm shift from CRM to a new paradigm of IF, leveraging the ubiquitous virtual presence of sales and customer engagement, which provides substantially richer capacity to access information.

We argue this new paradigm solves major problems of the current CRM paradigm. First, the advantage of the IF paradigm starts in automating knowledge production. In the CRM paradigm, only human produce knowledge: sellers manually update CRM to record a fraction of their knowledge in their mind. Data and knowledge in CRM is incomplete, inconsistent and outdated. In the new paradigm, the IF system directly perceives sales and customer engagement through API access to salespeople's email mailbox, calendar, telephony and video conferencing software, then automates comprehension and projection to produce consistent and up-to-date knowledge. The IF approach alleviates costly manual data, and increases objectivity in knowledge production by removing human biases and manipulation. Second, the IF paradigm enables more systematic knowledge production which not only serves specific use cases of teams who directly produce the knowledge, but also enables other teams or departments to leverage them. This enhances individual and team SA and reduces information silos by increasing knowledge transferability. Third, the IF paradigm can also enable more powerful ecosystem of applications by sharing or ingesting computable knowledge assets with developers of ecosystem partners.

## II. SITUATION AWARENESS IN SALES AND CUSTOMER ENGAGEMENT

### A. Overview of Sales

There are two major categories of sales based on the types of customers: Business-to-Consumer (B2C) and Business-to-Business (B2B). About 4.5 million people work in B2B sales in the U.S., as estimated by Forrester Research [12]. B2B sales often has longer sales cycles, more complex customer problems and solutions, more rational and rigorous buyer decision making processes, and greater number of decision makers on the buying side [13]. Sales can be categorized into three major phases: prospecting, closing, and post-sales. *Prospecting* is to identify potential customers (i.e., prospects), to engage with them for information discovery and determining their qualification, and to get their further time commitment. In *Closing*, salespeople engage with qualified prospects for in-depth selling activities (e.g., demo, trial or pilot) and evaluations to establish the value of the product or service to the customers' problem and need and to close the deal. *Post-sales* refers to account management (e.g., renewal, cross-sell, and upsell) and customer success management (e.g., customer training and business reviews); it becomes more and more important with the booming of subscription business models [14].

In a sales team, there are different roles: sales representatives (reps), sales managers, and sales operations. Sales reps engage with customers to drive sales and are responsible for delivering their sales quotas. Sales managers are responsible for building a team of high-performing sales reps and accountable for the team's sales quota. They regularly conduct deal reviews and coaching with sales reps, to understand the health of the deal pipeline, to identify deal risks and skill gaps and provide guidance. Sale operation conducts sales planning and quotas setting, forecasting and performance metrics analysis, sales process optimization, as well as assessment and adoption of software systems.

B2B sales is fundamentally a team sport, where salespeople with different specializations and seniority levels collaborate closely to navigate the buyer's organizational structure aligning multiple stakeholders. Fig. 1 illustrates a deal unit in B2B sales, which consists of heterogeneous roles of players in the selling and buying sides.

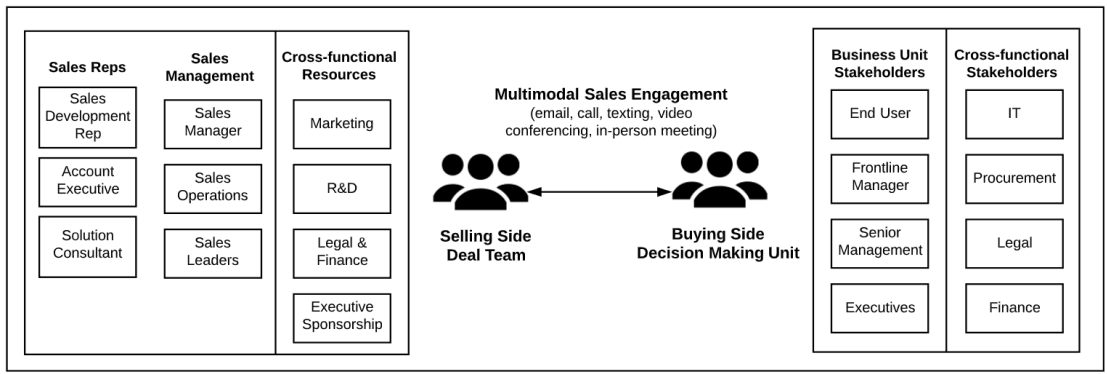

Fig. 1. A Deal Unit of Selling-Buying Sides in B2B Sales

### B. Individual and Team Situation Awareness in B2B Sales

We argue that individual and team SA is at the center of effective sales and customer engagement, which is a gold mine not yet explored by researchers in human factors and information fusion communities. In a team environment, where multiple people are collaborating to achieve goals, team SA is defined as "the shared understanding of a situation among team members at one point in time" which requires communications at each level of perception, comprehension and projection, and which depends on team processes that facilitate communications [15].

*1) Individual SA of sales reps*: Sales reps need to perceive various signals and data in the environment through active discovery, such as extracting information and clues from customer-facing conversations, researching the buyer's business or competitors' campaigns. Comprehension of situations such as customer problem and competition are crucial for sellers to articulate and establish their value and differentiation. Projection of future states and events is crucial for sales reps to decide when to take what actions and adapt quickly and fluently to customer reactions as well as other changing conditions in the environment.

*2) Individual SA of sales managers and sales operations*: SA is the foundation for sales managers to coach each sales rep in the team, and oversee the portfolio of deals owned by the team; it is also the key for sales operations to conduct revenue forecasting and diagnose how to improve the sales process and sales methodology. Sales managers and sales operations need to indirectly perceive and comprehend what is going on in sales engagement through sales reps' information and knowledge sharing.

*3) Team SA of the selling side*: Team SA is the key for effective teamwork, which is prevalent in B2B sales. There are multiple sales reps interacting with the buying side throughout the buying journey. In addition, given the role specializations, often sales reps with different roles are responsible for different phases of the sales processing (i.e., prospecting, closing, and post-sales) and there are hand-offs in each phase transition.

*4) Individual and Team SA of the buying side*: Buying complex products or solutions, such as enterprise software or manufacturing solutions, is often difficult, time-consuming, stressful or even frustrating [16]. SA is the key for B2B purchase decision making, which involves perceiving relevant information and signals in the environment, comprehending the problem to be solved and each solution being evaluated, and projecting the viability, return on investment (ROI), and risk factors of each solution.

*C. Computable Representation of SA*

Fig. 2 illustrates Endsley's Model of SA and corresponding computable representations of SA as knowledge assets.

*1)* **Signals for perception:** A signal is an inferred or derived property of an entity. A signal contains useful information about the state of the environment. As we develop a new signal, we increase our capability to observe what is going on. Example: prospect personality, intent of prospect reply email.

*2)* **Situations for comprehension**: A situation is a summarized representation or comprehension of environment system, based on which an agent decides and performs actions. A situation is derived from signals and attributes of all relevant entities. Fine-grained situation comprehension is required for well-informed and mindful decisions and actions. Example: A computable summary of a deal's situation.

*3)* **Patterns for projection:** A pattern is a projection into the future state based on the situation and possible actions. Two types of patterns include: (1) predicting the probability of successful outcome given the situation, and (2) predicting probability of a successful outcome given possible actions and the situation. A pattern captures correlational (Machine Learning) or causal relationship (Experimentation) in historical data. Example: The predicted probability that a deal can be won at each situation.

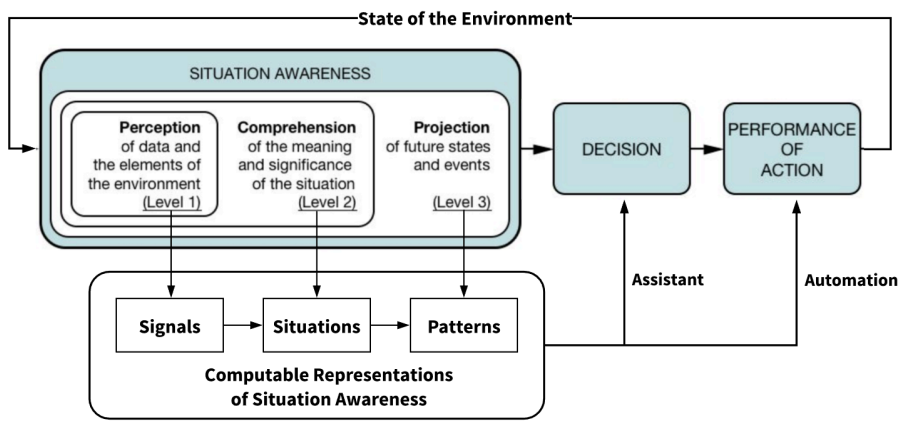

Fig. 2. Endsley's Model of SA and Computable Representations of SA[1]

### III. A VISION OF PARADIGM SHIFT: FROM CRM TO INFORMATION FUSION

*A. Unsolved Problems in the Paradigm of CRM*

CRM systems are "information systems that enable organizations to contact customers, provide services for them, collect and store customer information and analyze that information to provide a comprehensive view of the customers" [17]. CRM is the dominant system of records today for sales and customer management, which serves as the information system for knowing "what is going on". There are three major unsolved problems in CRM paradigm.

*1) CRM paradigm depends on costly manual data entry and updating for knowledge production, which leads to highly incomplete and outdated data and knowledge in CRM and less time selling.* In the CRM paradigm, manual data entry by sales reps is the primary source of knowledge production. Sales reps manually update CRM to record only a fraction of their SA in their mind. It is a well-recognized that sales reps do not want to spend their time manually updating CRM. According to a recent global survey studies, sales reps only spend 34% of their time actually selling, while spending 8% of their time on manual data entry and 9% time on other administrative tasks, and they list "inputting sales data and customer notes" and "logging activities" as two of the top 5 things they spent too much time on [4]. In fact, as estimated by Salesforce, the global leader of CRM market, 91% of CRM data is incomplete and 70% of that data decays annually [18].

*2) CRM paradigm falls short in solving the problems caused by human biases and manipulation in data entry and updating and by salespeople "gaming the system".* Relying on manual data entry and updating introduces human biases and strategic manipulation, which leads to untrustworthy CRM data. For example, "Sandbagging" is a widely recognized phenomenon that salespeople game the system by making intentionally overly conservative forecasts and by not updating CRM in a timely manner to hide progresses they made [19]. Research shows that salespeople of enterprise software are adept at gaming the timing of deal closure as well as offering aggressively lower price to take advantage of incentive compensation schemes, which can result in mispricing, therefore costing a significant proportion (6%-8% in the study) of revenue [20].

*3) CRM paradigm is incapable of producing high-level customer knowledge for facilitating team SA, which leads to frictions in knowledge sharing, information silos, and negative customer experience.* As noted by information systems researchers and practitioners, CRM is "mainly used to collect and organize customer information and not capable of generating high-level customer knowledge" and "organizations

---

[1] The illustration of Endsley's model of SA is adapted from the drawing by Dr. Peter Lankton [36], which was synthesized from [37] and [38].

often do not make good use of their CRM systems' capability to obtain knowledge from their customers" [17]. In CRM paradigm, knowledge is transferred mostly through talking to the right people. Certain amount of information and knowledge is gained at each stage in the sales process by each player, but it is often heavily siloed. Customers are often getting flooded with requests and touches from across the selling side, which leads to negative experience.

*4) The CRM paradigm is incapable of supporting the next generation of sales process optimization and SA enhancement, because of its incapability of producing computable knowledge assets representing SA.* The holy grail of sales management is to develop repeatable, scalable and continuously improving sales processes, methodology and enablement for sales reps, so that the sales team functions as a predictable revenue generating machine for the business. To achieve this ultimate goal, it is necessary to have a solid foundation of data and knowledge representing "what happened" and "what is going on". In other words, we need computable knowledge assets capturing situation awareness. The CRM paradigm becomes a bottleneck for the next generation of sales management advancement.

### B. A Vision on the New Paradigm of Information Fusion for Sales and Customer Engagement

We propose a vision of new paradigm of Information Fusion (IF) for sales and customer engagement, to overcome the unsolved problems in the CRM paradigm.

In the current JDL/DFIG (Joint Director Laboratories/Data Fusion Information Group) model, there are six levels [11], [21], [22]:

- Level 0 – Data assessment: estimation and prediction of states of sub-object entities (e.g., signals, features).
- Level 1 – Object assessment: estimation and prediction of states of entities on the basis of data association and processing.
- Level 2 – Situation assessment: estimation and prediction of relations among entities, which may require adequate user input.
- Level 3 – Threat/impact assessment: estimation and prediction of effects on situations of planned or estimated actions.
- Level 4 – Process refinement: adaptive data acquisition and processing to support sending objectives.
- Level 5 – User refinement, and Level 6 – Mission refinement.

The key in the new paradigm is to automate the production of computable knowledge assets by extracting signals directly from sales engagements, updating situations, and mining patterns. There are several major advantages of the IF paradigm which solves the problems in CRM paradigm.

*1) Scalability through Automated Knowledge Production:* IF paradigm automates the production of computable knowledge assets. This reduces the burden of manual data entry and enables more consistently captured and up-to-date data and knowledge. This also enables more scalable knowledge production through automation.

*2) Objectivity through Removing Human Biases*: The IF paradigm improves objectivity of knowledge production by removing potential human biases and strategic manipulation in sellers' manual data entry or updates. However, the IF paradigm can introduce algorithmic biases, which exist in many autonomous systems and requires deliberate efforts in identification and intervention [23].

*3) Cross-departmental Applicability through Decoupling Knowledge Production from Specific Needs of Its Producers:* The IF paradigm enables more systematic knowledge production which not only serves specific use cases of teams who directly produce the knowledge, but also enabling other departments to leverage them.

*4) Enhanced Transferability of Knowledge through Computable Knowledge Assets*: The IF paradigm enhances individual and team SA and reduces information silos by increasing knowledge transferability. It can also enable more powerful ecosystem of applications by sharing or ingesting computable knowledge assets with developers of ecosystem partners.

Table 1 summarizes the comparison between the two paradigms.

TABLE 1. Comparison of Two Paradigms from Knowledge Aspects: CRM and IF

| Aspects | CRM Paradigm | IF Paradigm |
|---|---|---|
| **Production of Knowledge** | **Environment → Sellers → CRM**<br><br>Only sellers can produce knowledge. Then sellers manually update CRM to record a fraction of their knowledge in their mind. This means data and knowledge in CRM is incomplete, inconsistent and outdated. | **Environment → IF**<br><br>Production of knowledge assets is automated by extracting signals directly from sales engagements, updating situations, and mining patterns. This enables consistent and up-to-date knowledge. In particular, Natural Language Processing (NLP) is leveraged to extract signals from sales conversations (emails, calls, texting, and meetings). |
| **Scalability of Knowledge** | **Better Sellers → Better Knowledge**<br><br>Given only sellers can directly produce knowledge, we need sellers who have higher Emotional Intelligence, more experienced and more mindful to produce better knowledge. And every seller need invest more time to better update their knowledge in CRM. | **Enhancement of IF → Better Knowledge**<br><br>Better knowledge is produced by developing more and higher quality signals, finer-grained situations, and mining more accurate and causally valid patterns. |
| **Objectivity** of Knowledge | **Human Biases and Strategic Manipulation** Relying on manual data entry and updating CRM introduces human biases and strategic manipulation. "Sandbagging" is a widely recognized phenomenon that salespeople game the system by making intentionally overly conservative forecasts and by not updating CRM timely to hide progresses they made. | **Objectivity Subject to Algorithmic Biases** IF paradigm automates the perception, comprehension and projection of environment, which improves the objectivity of knowledge produced by removing human biases and strategic manipulation. However, the IF paradigm can introduce algorithmic biases, which exist in many autonomous systems and requires deliberate efforts in identification and intervention. |
| **Applicability** of Knowledge | **Departmental Applicability** Knowledge captured by sellers in CRM is mostly intended to serve the specific operational and managerial needs in sales department, rather than serving broader and diverse needs of the entire company. Sellers' knowledge about potential buyers and customers are not utilized by other departments such as product and engineering to fundamentally improve product offerings. | **Cross-departmental Applicability** Voices from potential buyers and customers, together with other signals from the environment, are systematically listened and processed to produce knowledge and insights for all relevant departments across the company. The IF paradigm enhances Team SA not only at department level but at the entire company level. |
| **Transferability** of Knowledge | **Sellers → Everyone else** Given there is not much captured in CRM, knowledge can only be transferred through talking to the right people who have the knowledge. This friction leads to information silos and compromised buyer experience. | **IF → Everyone else + Ecosystem** The IF paradigm produces computable knowledge assets, which can be delivered in easy-to-digest format to people in the sales team, thereby removing information silos. Moreover, it can share or ingest computable knowledge assets with developers and independent software vendors (ISVs), to create a more powerful ecosystem than what can be achieved by a CRM. |

*C. Why now? Timing and Market Conditions for the Paradigm Shift.*

We argue that now is the perfect timing for the paradigm shift, from both the technical feasibility side and the business demand side.

*On the technical feasibility side, the ubiquitous virtual presence of sales and customer engagement and the rapid growth of inside sales enables substantially richer capacity to access information in this domain.* In the past decade, inside sales (i.e., remote sales) has been growing dramatically, far outpacing that of field sales (or outside sales); not only inside sales reps are selling remotely, it is estimated that even field sales reps interact with customers remotely more than 50% of their time [5], [24]. The increasing virtual presence of sales and customer engagement leads to higher quantity and quality of "sensors" for information access. By obtaining authentication from sales reps, the IF system can leverage the API access provided by office productivity software suites (e.g., email, calendar, telephony systems, and video conferencing tools) to observe sales reps' activities and communications, including both internal activities and customer-facing engagement. In fact, this pattern of accessing sales reps' emails and calendars has become common practice by sales technology vendors [25], [26], although most of the use cases are no more than low-level information fusions (i.e., level 0 and 1 in JDL/DFIG model).

*On the business demand side, there is tremendous opportunities and buzz about leveraging artificial intelligence (AI) to transform business in general and sales and customer engagement in particular.* However, solid AI applications cannot be built on shaky data and knowledge foundations. As recognized by domain experts [6], integrating data from diverse sources is a major challenge for realizing the potential of AI to transform selling: *"Assembling the data for a one-time use is difficult enough. Creating the processes needed to continually refresh the data can be daunting, time consuming and expensive."* We believe that the old paradigm of CRM is fundamentally incapable for providing a sustainable foundation of data and knowledge for the AI transformation. Companies will realize that the long-term success and viability of AI transformation depends on whether they can master the information fusion challenges. In addition, they will realize this after their initial rounds of investment in AI applications which show limited business impact and unsatisfactory return-on-investment (ROI) due to the bottlenecks in the old paradigm of CRM summarized in Table 1.

IV. A PATH TO INFORMATION FUSION IN SALES AND CUSTOMER ENGAGEMENT

We propose a path for developing IF system in sales and customer engagement application domain, describing core entities, and the key aspects in low-level and high-level information fusion.

*A. Core Categories of Entities*

There are 7 major categories of entities in B2B sales and customer engagement.

- **Accounts**: Buying side companies.
- **Contacts**: People on the buying side.
- **Employees**: People on the selling side, such as sales reps, sales managers, and collaborators in cross-functional teams.
- **Engagements**: Customer-facing interactions between the selling and buying sides, including emails, calls, meetings and other social channels like LinkedIn. Engagements are at the center of the graph of linked entities because they are the major driver for changes in the state of the environment system.

- **Projects**: Projects are entities capturing a series of efforts and activities to achieve a certain goal. In the sales and customer engagement domain, projects include marketing campaigns, prospecting project, closing project, and post-sales project. In addition, there can be hand-offs between projects.
- **Content**: Customer-facing content such as email template, call scripts, and documents like case studies and white papers.
- **Orchestration**: Orchestration entities are workflow templates consisting of multiple steps of engagements (i.e., customer touchpoints) and logical control flow of these steps (e.g., time interval between two steps and conditions for ending). Orchestration entities codify standardized practices of sales engagement into workflow and can further enable experimentation (e.g., AB testing of two different ways of orchestration).
- **Competitors:** Competitive context is a key part in SA for sales. One of the biggest shortcomings of CRM is the lack of understanding of competitors' advantages and disadvantages and their influence on prices and optimal offers and product specification.

B. *Low-level Information Fusion (Level 0 and 1)*

- **Data access and assessment leveraging the ubiquitous virtual presence of sales and customer engagement:** This requires the IF system to get authentication from sales reps, obtaining API access to their email mailbox, calendar, telephony and video conferencing software, so that the IF system can directly observe and assess both internal activities and customer-facing engagement.
- **Data enrichment**: Data enrichment can be based on 1st party data or 3rd party sales intelligence providers. One important use case is to enrich attributes of new contacts (e.g., name, job title, seniority) involved in emails or calendar meetings based on their email addresses.
- **Signals enrichment using NLP**: Conversations in natural language is the key format of sales and customer engagement. An important element of IF is to leverage NLP to extract signals from multimodal sales conversations in emails, texting, calls, and meetings.

Low-level fusion itself can improve SA of sales reps. For example, we developed intelligent automation capabilities at Outreach, a leading sales engagement platform, to detect whether an email reply from a prospect is an out-of-office automatic reply, and then parse the prospect's return date and new contact information mentioned there. The extracted information is leveraged to improve buyer experience and seller effectiveness by pausing sales communication until the prospect returns [27].

C. *High-level Information Fusion (Level 2 and 3)*

*1) Estimation and prediction of the relations between Engagements and Projects is the foundational task in Level 2 Fusion.* There are two aspects. First, when the IF system observes engagement activities (e.g., email, call, meeting), it needs to attribute and link each engagement to its corresponding project, which is not trivial when multiple ongoing projects exist between the selling and buying side. Second, the IF system needs to estimate and predict the relations between the engagement and the corresponding project, enabling comprehension about what is the project situation when the engagement took place and what is the goal of the engagement with respect to the project.

*2) Assessing the situation of Projects is the key goal in Level 2 Fusion.* The Projects entity captures the basic units of sales and customer engagement. A project is a purposeful initiative to achieve a certain goal such as prospecting to book meeting, closing a deal, renewing an existing customer, or expanding an existing customer. Assessing the situation of projects require leveraging signals of all other entities and their estimated relations. The computable representation of situations needs to be designed based on what is the decision to make or action to perform.

*3) Estimation and prediction of the probability of project success and progression conditional on project situation and planned action (situation-action pair) is the key goal in Level 3 Fusion.* This leads to computable knowledge assets of patterns, which can be correlational or causal relationships.

Fig. 3 illustrates core categories of entities and Level 0 to Level 2 fusion in sales and customer engagement. The dotted lines indicate relationships that do not exist as a result of incomplete manual data entry in the current CRM paradigm. Establishing these relationships through estimation and prediction from raw data streams is the foundation in the IF paradigm.

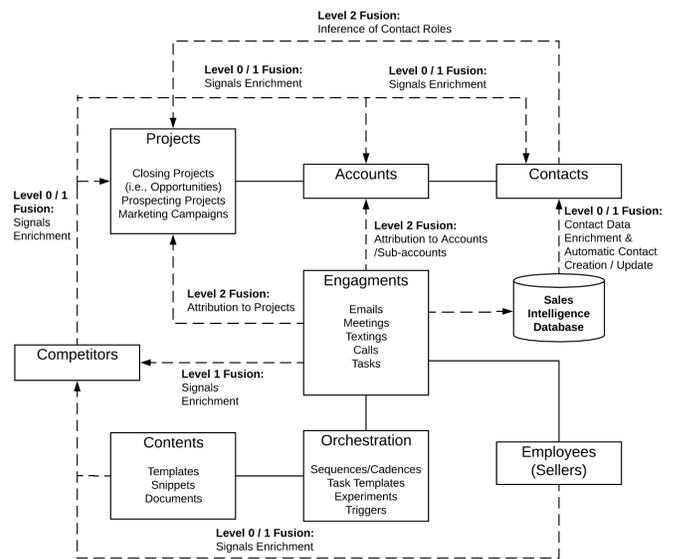

Fig. 3. Illustration of Core Categories of Entities and Level 0 – Level 2 Fusion in B2B Sales

## D. Process Refinement (Level 4)

Level 4 fusion is considered as a meta-process, a process that monitors and optimizes the overall IF process, as well as process refinement and resources allocation optimization for better achieving mission goals [28]. In the application domain of sales and customer engagement, Level 4 fusion has two aspects: refinement of the IF process, and refinement of the sales process and organizational design.

*Refinement of IF process.* First, measuring the peformance both at the overall system level and at the component level to identify refinement opportunities. Completeness, correctness, and computational performance are key dimensions of performance. Second, expanding the capabilities for IF system to directly observe the environment is the key for substantial enhancement – digitizing the part of sales and customer engagement that are not virtually presented yet. This does not limit to the customer-facing engagement but also the internal business process for sales team to coordinate, manage, and operate. Third, human-in-the-loop feedback from users and from human annotators are scarce resources that require optimization. Low-level fusions involve developing, maintaining, and improving NLP and machine learning capabilities, where efficient utilization of human annotators resources is often challenging in both scientific and operational aspects. High-level fusions often require feedback from users with subject matter expertise.

*Refinement of the sales process and organizational design.* First, improving the sales process based on data and facts is a key to realize the value of IF systems. With enhanced SA and deeper understandng of sales processes backed by data, sales team can go beyond their static and rigid playbook to better adapt the dynamically evolving situations of a deal. Second, greater SA also allows better human resources investment and allocation, as well as incentive compensation. With better understanding of the contribution of each cross-functional roles in different situations, companies can be more confident in investing supporting resources. Similarly, design and implementation of incentive compensation can benefit from better evaluation of each person's contribution in closing a deal. Third, refining competitive strategy and pricing strategy leads to strength in a dynamic market environment. Enhanced SA has the potential to enable significantly more nuanced and effective pricing as well as increased sales. The enhancement in pricing alone may add 5% to 10% to the bottom line.

## E. Privacy, Compliance and Governance

The significance of Privacy, Compliance and Governance is of first-order for developing IF systems for sales and customer engagement. The emergence of new strict data privacy regulations such as GDPR (General Data Protection Regulation) and CCPA (California Consumer Privacy Act) further reinforces the principle that privacy, compliance and governance must to be considered as a first-class citizen from the start rather than being thrown in as an afterthought.

How to architect decision support systems for GDPR/CCPA compliance is still a research frontier. We highlight some important aspects in the context of our application domain, without attempting to be comprehensive. First, access privileges to salespeople's email mailbox, calendar telephony and video conferencing softwares need to be minimized following the *purpose limitation principle* and transparently communicated when seeking user consent. Second, detection and anonymization of sensitive personal information, in both structured data and unstructured data, shall be implemented at the data acquisition layer, so that they are never stored in the IF systems. To achieve this with high accuracy is itself a nontrivial NLP challenge. Third, data provenance needs to be consistently tracked to enable the *right to be forgotten* of prospects and customers. There is an active research area on how to do this in a reliable and scalable way leveraging technologies of semantic web and process mining [29], [30].

## F. Review of Example Applications

In this section, we review some example applications.

The first example application area is **automated sales leads acquisition and qualification**. It targets pain points in the beginning of Prospecting phase: as a sales rep, how can I acquire high quality sales leads, and how can I rank the sales leads according to their qualification so that I can prioritize my time on those more qualified? Solutions typically involve crawling the web to harvest more sales leads and to enrich the information of existing leads, then predict the probability that the lead can be converted if a sales rep contacts the lead [31]–[33]. This application area, however, is limited to the early phase in the sales process, before salespeople start engaging prospects or customers. Therefore, it does not tackle the challenges of achieving SA in the more dynamic and complex environment in sales and customer engagement.

GE Capital's financing lead triggers system [31] is worth special discussion for two reasons: (1) it is an industry-grade application deployed to hundreds of sales reps with documented business impacts, (2) its lead qualification is based on dynamic patterns in financial time series data, rather than only based on static firmographic information, which makes its triggers time sensitive and thus more actionable for sales reps to identify companies who are actively in need of financing. From the IF paradigm perspective, the system automates *perception* of the environment from signals of various financial time series data of target companies, *comprehension* of situations by summarizing the signals into representations of financial metrics, and *projection* of whether a company would have financing transactions in the next 6-12 months based on patterns learned from historical data or defined by experts.

The second example application area is **sales process discovery and coaching**. It targets pain points of sales managers: they lack of SA of actual sales processes for coaching sales reps to do better. At Outreach, we developed an objection handling coaching system. The system automates *perception* of prospect reply intent extracted using NLP. Specifically, we leverage transfer learning techniques to develop intent classification model to predict whether an email reply from a prospect is positive, objection, or unsubscription intent, as well as finer-grained subclasses such as "objection – no budget" or "positive – willing to meet" [34], [35]. It also *comprehends* of situations of whether a prospect's initial reply is a positive or objection intent and the sales rep's behavior in following up. Then it *projects* of whether a prospect can be converted into a deal conditional on the prospect's reply intent and the rep behavior in following up.

In a client pilot study, we discovered that the probability of conversion conditional on initial objection intent is about one third of that conditional on initial positive intent, which is much higher than our client expected. We discovered that their top performing sales reps responded to objection replies as consistently and promptly as they did for positive replies, while the vast majority of reps did not. The sales management was surprised by this result, as they believed that objection replies from prospects were doomed thus explicitly advised their reps to prioritize on responding to positive replies. Our results show that their best sales reps were actually not following the guideline and benefited from doing so. Our client changed their sales process guideline to require and to measure sales reps' objection handling response rate and response time, and they estimated a sizable increase in their pipeline generation by this process refinement.

## V. Conclusions, Challenges, and Future Work

In this paper, we propose a vision and path for the paradigm shift from CRM to the new paradigm of IF, in the application domain of sales and customer engagement. Our core contribution is in developing the framework for applying the IF approach in this emerging application domain, and articulating the paradigm shift opportunity it brings about.

For realizing the promise of this paradigm shift, we need breakthroughs in developing killer applications that can solve a subset of customer pain points thoroughly. Through developing these killer applications, we will deepen our understanding and develop new techniques for further driving through the paradigm shift. In addition, measuring individual and team SA in sales and customer engagement, and establishing the relationship between SA and outcomes (e.g., sales quota attainment) is a foundational area for future research.

## Acknowledgment

The author would like to thank Pavel Dmitriev, Rajiv Garg, Amritansh Raghav, Abhishek Abhishek, Eugene Ho, Sarah Phillips, and Ben Edick who provided very valuable discussions. The author would like to thank Preston McAfee, Pavel Dmitriev, Manny Medina, and Jim Lee for insightful comments and feedback for an earlier draft. All errors remain the author's own.